\documentclass[11pt]{article}
\usepackage{pgfplots}
\usepackage{subcaption}

\usepackage{amsmath,amsfonts,bm}

\def\eqref#1{equation~\ref{#1}}
\def\1{\bm{1}}

\DeclareMathAlphabet{\mathsfit}{\encodingdefault}{\sfdefault}{m}{sl}
\SetMathAlphabet{\mathsfit}{bold}{\encodingdefault}{\sfdefault}{bx}{n}

\DeclareMathOperator*{\argmax}{arg\,max}

\usepackage{hyperref}
\usepackage{url}
\usepackage{adjustbox}
\usepackage{fullpage}

\title{Detecting malicious PDF using CNN}

\author{%
  Raphael Fettaya\\
  Tel Aviv University\\
  \texttt{raphaelfettaya@gmail.com} \\
  \and
  Yishay Mansour  \\
  Tel Aviv University \\
  \texttt{mansour.yishay@gmail.com} \\
\
}

\begin{document}

\maketitle

\begin{abstract}
Malicious PDF files represent one of the biggest threats to computer security. To detect them, significant research has been done using handwritten signatures or machine learning based on manual feature extraction. Those approaches are both time-consuming, require significant prior knowledge and the list of features has to be updated with each newly discovered vulnerability. In this work, we propose a novel algorithm that uses an ensemble of Convolutional Neural Network  (CNN) on the byte level of the file, without any handcrafted features. We show, using a data set of 90000 files downloadable online, that our approach maintains a high detection rate (94\%) of PDF malware and even detects new malicious files, still undetected by most antiviruses. 
Using automatically generated features from our CNN network, and applying a clustering algorithm, we also obtain high similarity between the antiviruses' labels and the resulting clusters. 
\end{abstract}

% Include the chapters of the thesis, as separate files
% Just uncomment the lines as you write the chapters

\section{Introduction}

\subsection{Malware in PDF}
Malware programs are still making newspapers' headlines. They are used by criminal organizations, governments, and industries to steal money, spy, or other unwanted activities.
As millions of new malicious samples are discovered every day, spotting them before they harm a computer or a network remains one of the most important challenges in cybersecurity. During the last two decades, hackers kept finding new attack vectors, giving malware multiple forms. Some use the macros in Microsoft Office documents while others exploit browser's vulnerabilities with javascript files. This diversity raises the need for new automated solutions.

Portable Document Format (PDF) is one of the most popular types of documents. Despite the lack of awareness of the population, it also became an important attack vector (AV) for computer systems. Dozens of vulnerabilities are discovered every year on Adobe Reader, the most popular software for reading PDF files \cite{cve}, allowing hackers to take control of the victim's computer.
PDF malware can be segmented into three main categories: (i) exploits, (ii) phishing, and (iii) misuse of PDF capabilities.
Exploits operate by taking advantage of a bug in the API of a PDF reader application, which allows the attacker to execute code on the victim's computer. This is usually done via JavaScript code embedded in the file.
In phishing attacks, the PDF itself does not have any malicious behavior but attempts to convince the user to click on a malicious link. Such campaigns have been discovered recently \cite{bes} and are, by nature, much harder to identify. 
The last category exploits some regular functionality of PDF files such as running a command or launching a file. 
All those attacks can lead to devastating consequences, such as downloading a malicious executable or stealing credentials from a website.

Regardless of recent work in machine learning for malware detection, antivirus companies are still largely focusing on handwritten signatures to detect malicious PDF. This not only requires significant human resources but is also rarely efficient at detecting unknown variants or zero-day attacks \cite{register}.  Another popular solution is the dynamic analysis by running the files in a controlled sandboxed environment \cite{supelec}. Such approaches increase significantly the chance of detecting new malware, but take much longer and require access to a sandbox virtual machine. They also still require a human to define the detection rules according to the file behavior.

\subsection{Classical antivirus models for PDF files}
Antivirus vendors use a few different approaches to detect malware in PDF:
\begin{itemize}

  \item \textbf{Signature-based detection} It is the most basic and common method used to identify malicious files \cite{comodo}. Security analyst manually inspects a malicious file and extract one or several patterns from the byte code, the "signatures", that they store in a database. When analyzing a new file, they try to match their code segments with the one in the database. If a match occurs, the file is blocked.     
  \item \textbf{Static Analysis} Another rudimentary technique commonly used by antivirus is static analysis. In consists of applying heuristic-based rules on the content of a file to find potentially malicious action. The easiest approach is the search for keywords like /JavaScript, /OpenAction, or /GoTo which are related to an action that can be harmful to the computer. In absence of those tags, an analyst can confidently say that the file is benign \cite{singh2020malware} (although some attacks are managing to inject javascript code without requiring a javascript tag). 
  \item \textbf{Dynamic Analysis} It is a more expensive but potentially stronger method for detecting malicious behavior. It consists of running the file in a controlled environment (sandbox) and evaluates and retrieve the API calls and the network activity produced by the possible malware. Then, a program can apply heuristics on top of the activity logs like connecting to a malicious website or launching a subprocess \cite{singh2020malware}.
\end{itemize}

\subsection{Contribution}

In this work, we are using an ensemble of Convolutional Neural Network (CNN) in order to detect any type of malicious PDF files. Without any preprocessing of the files, our classifier succeeds to detect 94\% of the malicious samples of our test set while keeping a False Positive Rate (FPR) at 0.5\%. Our classifier outperforms most of the antiviruses (AV) vendors available in the {\it VirusTotal} website.% \newline
We also show that our CNN can successfully group more than 75\% of the malware into different families. Finally, we will present some examples on which we were able to detect an attack before the AV (zero-day).% \newline

To the best of our knowledge, this is the first paper using Neural Network to classify PDF malware. It is also the first one that investigates the ability to automatically classify malicious PDF into different families. 
Finally, as an attempt to build a baseline for detecting PDF malware, we open-sourced the list of the files used for the research. They are all downloadable from VirusTotal.

\noindent{\bf Paper organization:}
We first present the related research in machine learning for detecting malicious PDF and the usage of Deep Learning applied to Malware Detection in executable files (Section \ref{sec:related-work}). We describe how we built our data set in Section  \ref{sec:data}, and describe our model in Section \ref{sec:model}. We show our results on the data sets in Section \ref{sec:results}. We investigate the capability of our network to differentiate between malware types in Section \ref{sec:differentiating}. Our conclusion is in Section \ref{sec:conclusion}. % Introduction

\section{Related work}
\label{sec:related-work}

\subsection{Machine Learning for detecting malicious PDF}

Signature-based detection used to be the standard in cybersecurity, and it is the preferred solution where researchers use signatures to identify malicious PDF \cite{rau}. However, with the fast increase of threats, the work required by handwritten rules increased significantly, and machine learning has been extensively used in the last decade to enrich detection capabilities. 

Munson and Cross \cite{mun} define a list of features extracted from both static and dynamic analysis of the PDF files. Those features are trying to catch potentially harmful action made by the PDF such as: launching a program, responding to an user action, or running javascript. Some other features describe the format of the file itself, as the presence of an xref table or the number of pages. They use a very small dataset of 89 malicious and 2677 benign samples.
Eventually, they train a decision tree classifier and detect around 60\% of the malware for a precision of 80\% on a 5-fold cross-validation.

Stavrou and Smutz \cite{smu}  focus their feature extraction on metadata and structure of the documents. Structural features are for example the number of instances of specific indicative string (/JS, /Font), or the position of some objects in the file. Metadata features reflect higher-level characteristics of the file. An example of these is whether there is a mismatch in the unique identifier (pdfid0) of the PDF. Using an even training set of 10000 files, an operational dataset of 100000 files (297 malware), and the 202 manually selected features they build a Random Forest classifier. In the train set, with a 10-fold cross-validation, results are bigger than 99\% without any False Positive, while in the test set, they achieve a perfect malware classification with only 0.2\% FPR. Despite those impressive results, two serious signs warn of severe overfitting of the model. First, using a 10-fold cross-validation with random sampling increase the chance of having similar files in the train and test set at each iteration. Another strong overfitting sign is the fact that the most important feature of the model is the count of the appearance of the string '/Font'. Although this feature is not indicative of a malicious behavior, it helps to identify similar files as its variance is very high.  

Together with static analysis features, Tzermias et al. \cite{tze} emulate the JavaScript code contained in the file, and managed to detect 89\% of the malware in their test dataset. It is worth noticing that, although this approach is robust to obfuscation, it takes 1.5s for their algorithm to run on a single PDF file and requires the usage of a VM.

More recently, Zhang \cite{zha} is using Multi-Layer Perceptron (MLP) on 48 manually selected features, and achieves a better detection than 8 antiviruses in the market, obtaining a detection rate of around 95\% at a 0.1\%FPR. He used a big dataset of more than 100000 files including 13000 malware.

A PDF reader needs to recognize a set of keywords (tags) in the content of the file to display links, open images, or execute actions.
Instead of using predefined features, Maiorca, Giacinto, and Corona \cite{mai} decided to generate features based on those keywords. To choose which of them to incorporate, they first split their dataset into benign and malicious samples, then, for each of those classes, they run a K-Means clustering algorithm with K=2 to split tags into a frequent and not frequent group. Finally, the researchers take the union of the frequent tags for both classes and use them as features. Using a training dataset of 12000 files, they end up with 168 features. Following that, they train a Random Forest classifier and obtain a detection 99.55\% at 0.2\% False Positive Rate on a test set of 9000 files. Although the results look very good, and similarly to all the previous papers, the data is split randomly and the test malware are not necessarily older than the training ones, which creates a high risk of overfitting. More importantly, most of the malicious files are using JavaScript of ActionScript exploits, and we could suspect that the success is explained by the detection of those two tags.

Just like in XML, PDF exhibits a hierarchical structure. In this work, Srndic and Laskov \cite{vsr} use an open-source parser to retrieve the tree-like structure of the file. Instead of a single tag, here all the tags forming the path from the root to a leaf of the tree are concatenated to make a feature. Then only paths appearing more than 1000 times in the dataset are kept and used to train a Decision Tree and a SVM model. The evaluation is made using a huge dataset of 660000 files, mostly taken from VirusTotal, including 120000 malware. Few experiments are made taking data from different periods. In a specific setup, the authors are training their algorithm using 4 weeks of VirusTotal data and evaluating on the following one, repeating this experiment 6 times. They overall get a detection rate of 87\% at a 0.1\% FPR.

We emphasize that all those approaches require prior domain knowledge as they are based either on manual feature selection or at least on the parsing of the PDF structure.  

More recently Jeong et al. \cite{jeong} attempted to detect malicious streams inside PDF files using CNN straight from the byte code. They use an embedding layer at the top of their networks to convert the first 1000 bytes of a stream into vectors. Then, they train several networks together with more standard Machine Learning models and compare them. They achieve a detection rate of 97\% with a precision of 99.7\%.  Unfortunately, the experiment raises several concerns. First, they used a small even dataset of 1978 streams in which all malicious streams had javascript embedded. This encourages the convolutional filters to look for a specific string indicating the presence of JavaScript. Also, the benchmark is only comparing the networks with other Machine Learning models with a byte-level and not tag-level feature extraction as proposed in most previous research.

\subsection{Deep Learning in malware detection}

In the last years, there have been several efforts in using Deep Learning for detecting malware in executable files (type {\tt exe}). 
 David and Netanyahu \cite{dav}, in DeepSign, are using a Deep Belief Network with Denoising Autoencoders to automatically generate signatures based on executables' behavior. To do so, they run the files on a sandbox, extract the API calls, and create 5000 one-hot encoded features out of them. The signatures consist of 30 features generated by the network that is used in the classifier and achieves a 98.6\% of accuracy on their dataset.
 
 Pascanu et al. \cite{pas}, starting from the API calls, are generating an embedding of the malware behavior using an Echo State Network and a Recurrent Neural Network. They train the network to predict the next API call and use the last hidden state as a feature for a classifier. They obtain a detection rate of 98.3\% with 0.1\% of FPR.

Based on static analysis of exe files, Saxe and Berlin \cite{sax} are using manually extracted features from static analysis of executables that they use in a four-layer perceptron model and detect 95\% of the malicious files in their dataset at 0.1\% FPR.

Finally, Raff et al.\cite{raf}, in Malconv, are using Convolutional Neural Network on the raw bytes to detect malicious executables. They are using an embedding layer to convert the first bytes of the files into a matrix that later goes through a CNN. Using a dataset of more than 500000 files taken from two different sources, they manage to get a 90\% balanced accuracy. It makes their results significantly worse than other approaches but no preprocessing is required on the data, and predictions are done very efficiently. They also show that adding more files to the training (2 million here) can improve the accuracy by 1.2\%.
 % Background Theory 

\section{Data sets}
\label{sec:data}

For our experiments, we use 18296 malicious PDF files and 70551 benign ones. The malicious samples are taken from VirusTotal \cite{vt}. They were uploaded on the website between the 5/20/2008 and the 10/11/2018. They were all detected by at least 10 antiviruses by 11/20/2018.
The benign files were also obtained using VirusTotal from the same time range, checking that none of the antiviruses detect them as malicious after a rescan in April 2020.
We also download a set of 9300 newer malicious files, following the same rule, to evaluate the degradation with the time. For this set, the files appeared on the website between the 2/15/2019 to the 3/15/2019.
Following the assumptions that the benign PDF evolve much slower than the malicious one, we did not record their date of upload of the benign files.

We partition our malicious and benign sets into training, validation, and test set. 
In order to make sure our research imitates as well as possible the real-word malware classification challenge, the sets were chronologically organized.
For the train set, we use 16605 malicious files, from the oldest one to the ones that appeared in 5/22/2018, and 63497 benign samples. The validation set helps us tuning the hyperparameters of our model and contains 460 malicious files  (until 09/15/2018) and 1409 benign. The test set contains the rest.

To compare our model to previous approaches, we also use the Contagio dataset \cite{contagio} which contains 9000 benign PDF files and 11155 malware. Following Maiorca et al. \cite{mai}, we use a train and test set of 6000 files each. We also leave 20\% of the test set as a validation for our algorithm. % Experimental Setup

\section{Neural network architecture}
\label{sec:model}

%[YM]
We design and present three different models.
The first one is a minimalist model, with a single convolutional layer, with a window size of 16 bytes, a stride of 4 bytes and 128 kernels. It is followed by a global max pool and finishing with a linear layer of size 128 and a sigmoidal gate, we refer to it as ModelA.
The second model, ModelB, is similar to ModelA, but has an additional fully connected layer of 128 $\times$ 128 weights, right after the global max pool. The additional fully connected layer allows the network to combine the filters together and giving him much more freedom, we denote it ModelB.
Finally, the last model is a deeper model, that is built out of three convolutional layers: the first one with a window size of 16, a stride of 4 and 20 kernels, then a second with the same window size and stride but 40 kernels, finally the third one has 80 kernels a window size of 4 and a stride of 2. All those layers are immediately followed by a batch-normalization. We then inserted one fully connected layer 80$\times$80 also with batch normalization and a linear layer. We call this network ModelC.
Both the convolutional and fully-connected layers are directly followed by a ReLU in all the networks.

In order to turn the files' byte code to input vectors, we proceed as follows. Initially, to turn them to a fixed size, we select only the first 200kB, if a file is smaller it will be zero-padded at the end. Then, each byte, which is a number between 0 and 255, is mapped to a 16-dimensional vector. To obtain the mapping of the byte $i$, $0\leq i \leq 255$ we simply compute the product $W^\top x_i$ where $x_i$ is a vector in $\mathbb{R}^{255}$ that is equals to 1 at index $i$ and 0 everywhere else, and $W \in \mathbb{R}^{16 \times 255}$ is the mapping (or embedding) matrix. This operation is actually the first layer of all our networks and the matrix $W$ is trained by back-propagation together with all the other parameters of our models. Such a process is commonly referred to as learning a task-specific embedding, as the matrix $W$ is trained with the network and thus, optimized for our classification problem.

The training has been done for 13 hours on average, on a single Nvidia GPU and we ran 5 epochs on the training data.

As a regularization technique, we use early-stopping, taking the checkpoints that were giving the best detection at a 1\% false positive rate on the validation set. For ModelB we use dropout with $p=0.25$ after the fully connected layer. We recall that in ModelC we use batch normalization after the convolutional layers and the first fully connected.

Due to its simplicity, ModelA cannot detect complex patterns from the data. It only creates filters for some strings, using the convolutional layer, look for them in the file, and combine them for classification. This is very generic and, in many ways, similar to what antivirus does with signature matching.\newline
ModelB has an additional fully-connected layer, it is able to combine the output of the filters together. Hence it can use multiple strings as signatures. \newline
ModelC has the ability to leverage its multiple convolutional layers to create complex signatures from different parts of the file and combine them together.

Our ModelA and B have essentially the following semantics. The network passes to the max pool layer a list of values indicating the presence of some strings in the file.
Those strings, that we will call signature by analogy with the signatures generated by the antivirus vendors, are used to perform the classification. The same signatures are also (potentially) helpful to distinguish different families of malware.
%[YM]   

\textit{The detailed architecture of those three networks can be found in Appendix \ref{appendix:networks}.} % Experiment 1

\section{Results}
\label{sec:results}

In this section, we will present the results in our dataset.
Following the standard malware detection denomination, we refer to malware as positive examples and benign files as negatives. A False Positive (FP) is a benign file that has been detected as malicious and a True Positive (TP) is a malicious file that has been correctly detected.
Similarly, a False Negative (FN) is a malicious file that has been detected as benign and a True Negative (TN) is a benign file that has been correctly detected.
We will mainly use two metrics: the detection rate or recall (D) $\frac{TP}{TP + FN}$ and the False Positive Rate (FPR) $\frac{FP}{FP + TN}$. 
In the first part, we will present the results obtained on the test set, then we will evaluate the degradation of the model on newer files, and we will finish by presenting some examples of (almost) zero-day detection.

\begin{figure}[t]
\centering
\caption{ROC curve on train and test set for the three models}
    \begin{subfigure}[b]{0.48\textwidth} 
        \centering \includegraphics[width=\textwidth]{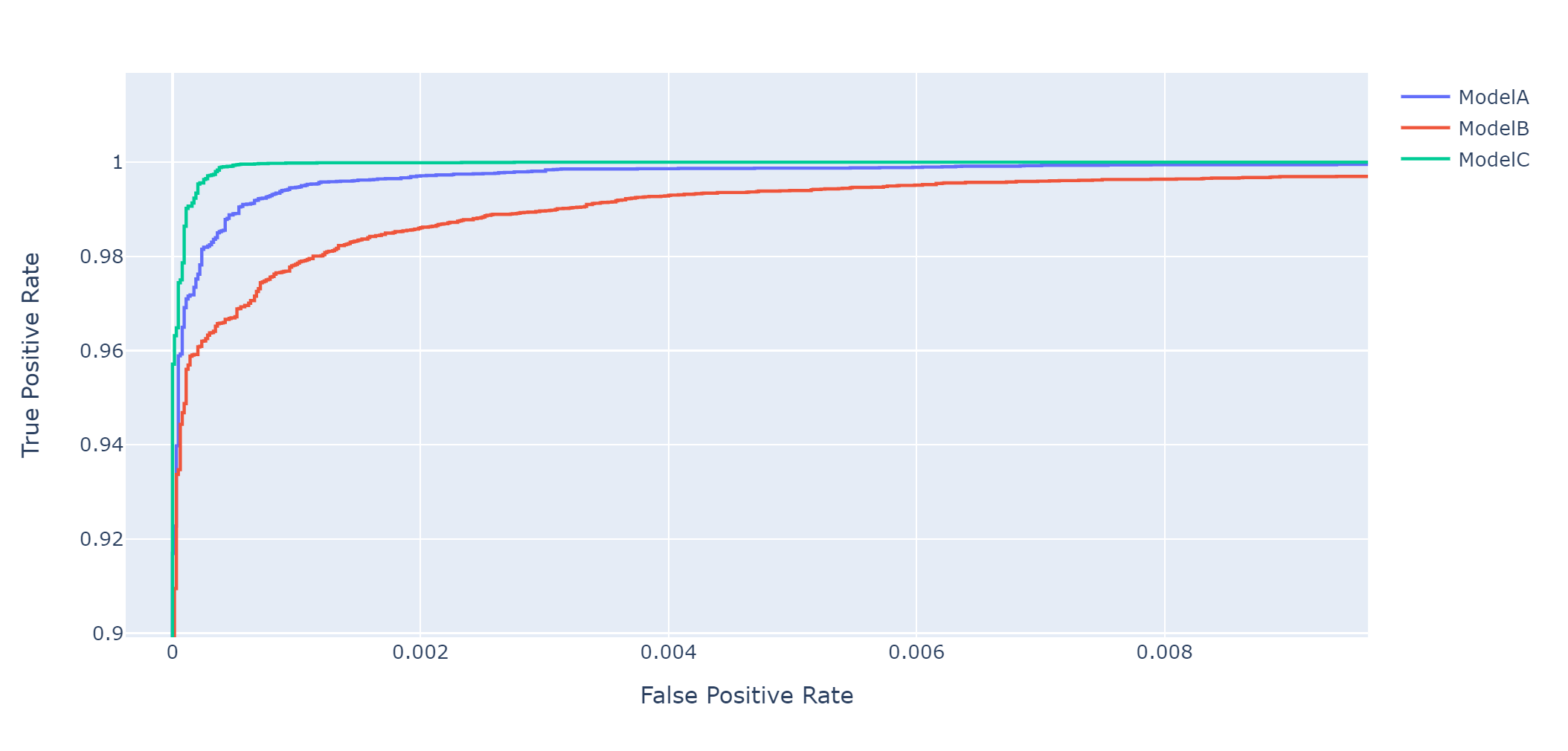}
        \caption{ROC on train set (from 0.9 TPR): despite its regularization ModelC is fitting better the training data, while dropout in ModelB reduces overfitting. %[YM: Adding focus]
        }
    \end{subfigure}
    ~ 
    \begin{subfigure}[b]{0.48\textwidth}
        \centering \includegraphics[width=\textwidth]{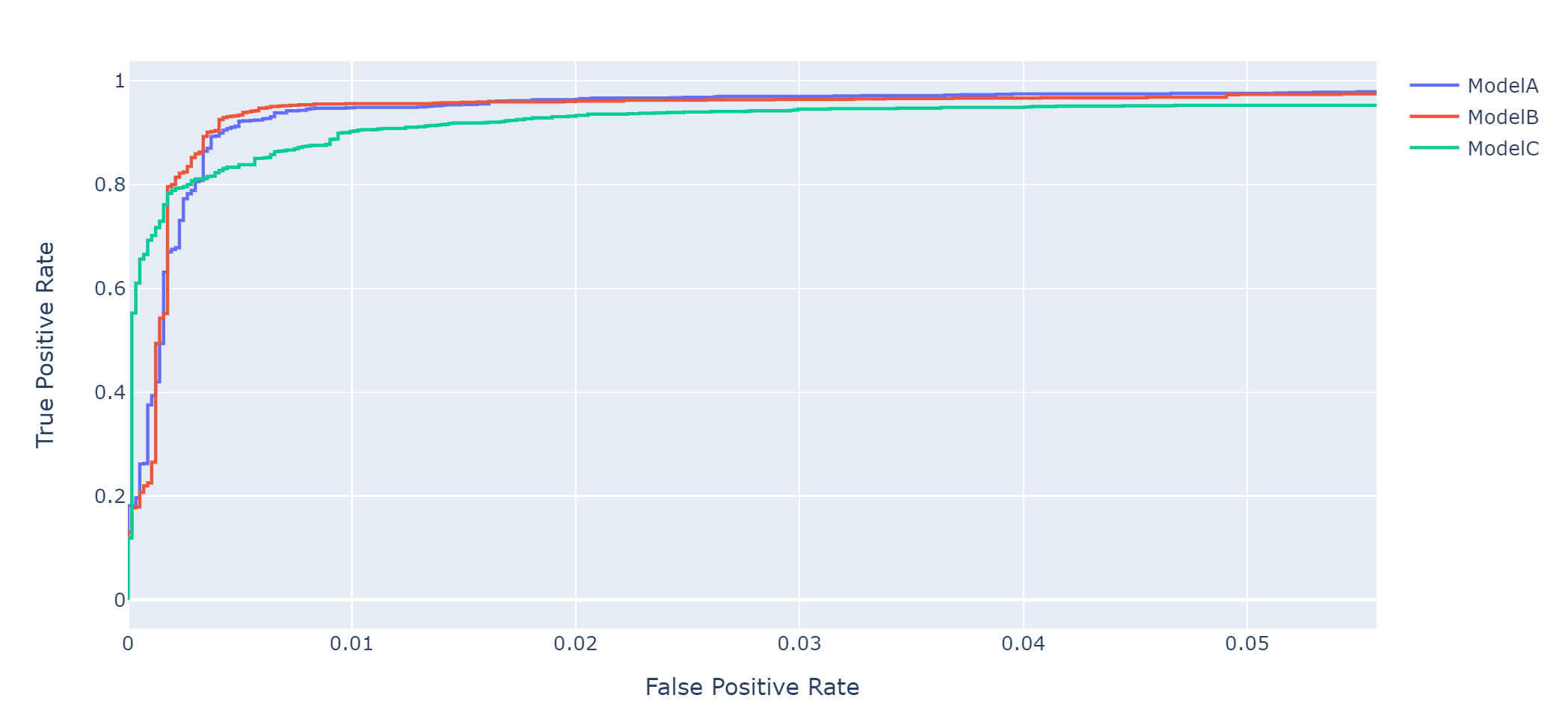}
        \caption{ROC on test set: ModelC is performing better on very low FPR whereas ModelB outperforms the two other around 1\%FPR.}
    \end{subfigure}
    \label{fig:roc1}
\end{figure}

\begin{figure}[h]
\centering
\caption{  Error bars for the detection rate of the models at 1\%, 0.5\% and 0.2\% FPR}
    \begin{subfigure}[b]{0.3\textwidth} 
        \centering \includegraphics[width=\textwidth]{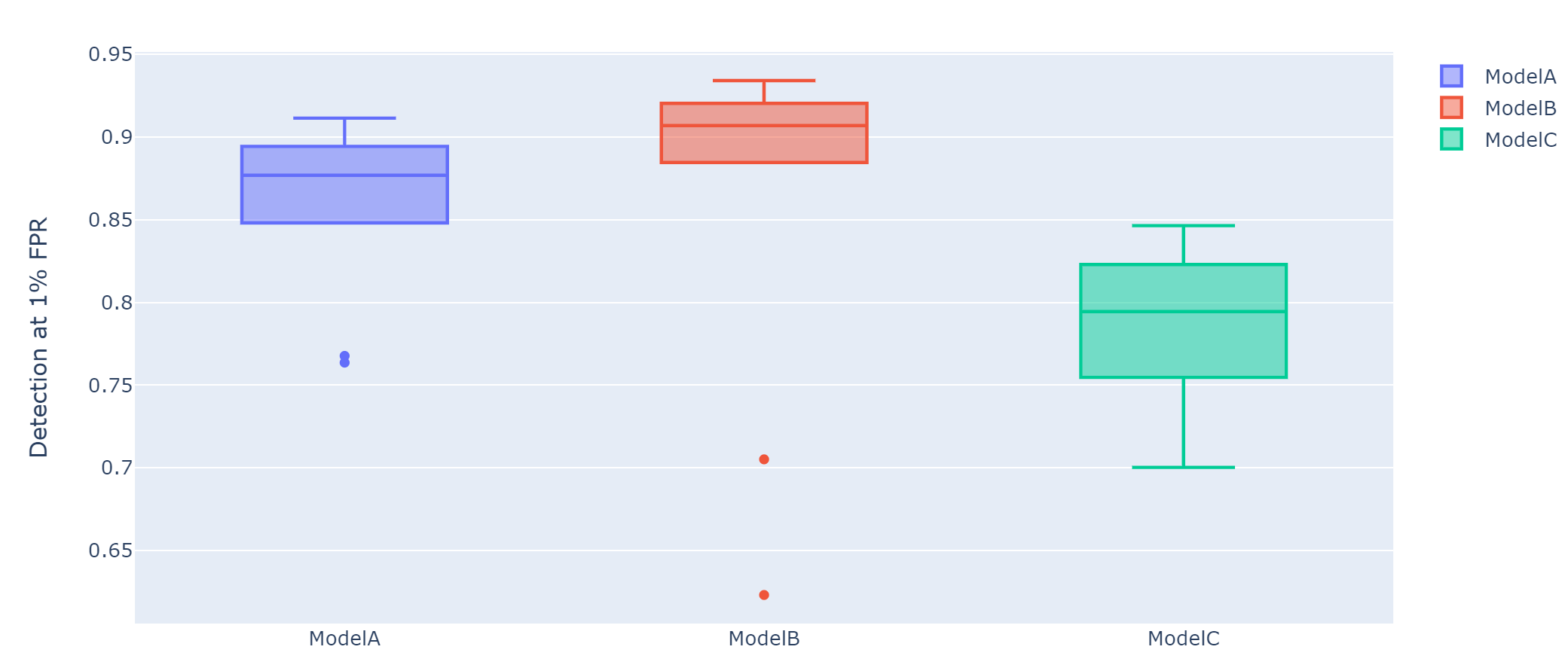}
        \caption{ Detection at 1\% per model. ModelB has the best detection of all the three models. As ModelC is doing  overfitting, results varies a lot between trainings. }
    \end{subfigure}
    ~ 
    \begin{subfigure}[b]{0.3\textwidth}
        \centering \includegraphics[width=\textwidth]{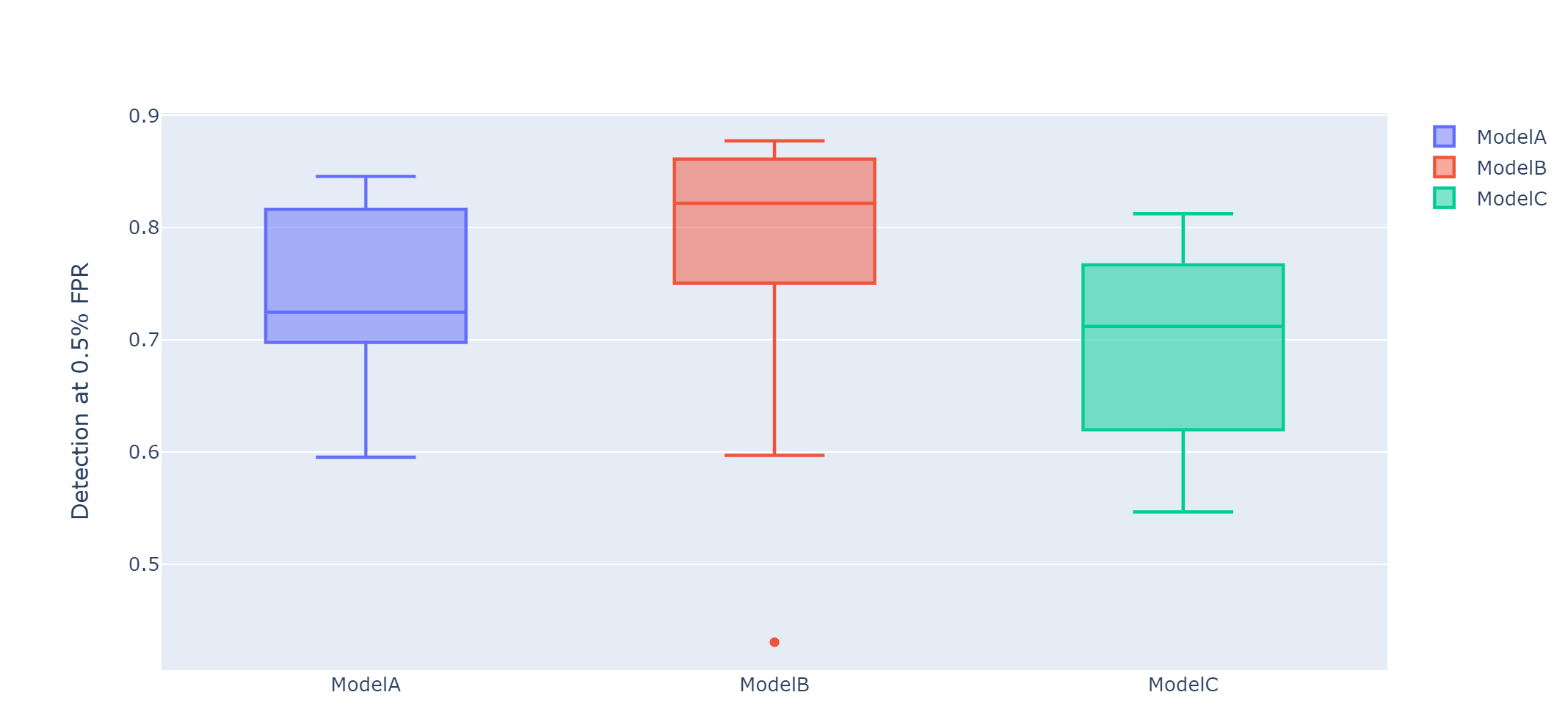}
        \caption{Detection at 0.5\% per model. While ModelB still performs better, the variance of the results of all the models increase drastically.}
    \end{subfigure}
    ~ 
    \begin{subfigure}[b]{0.3\textwidth} 
        \centering \includegraphics[width=\textwidth]{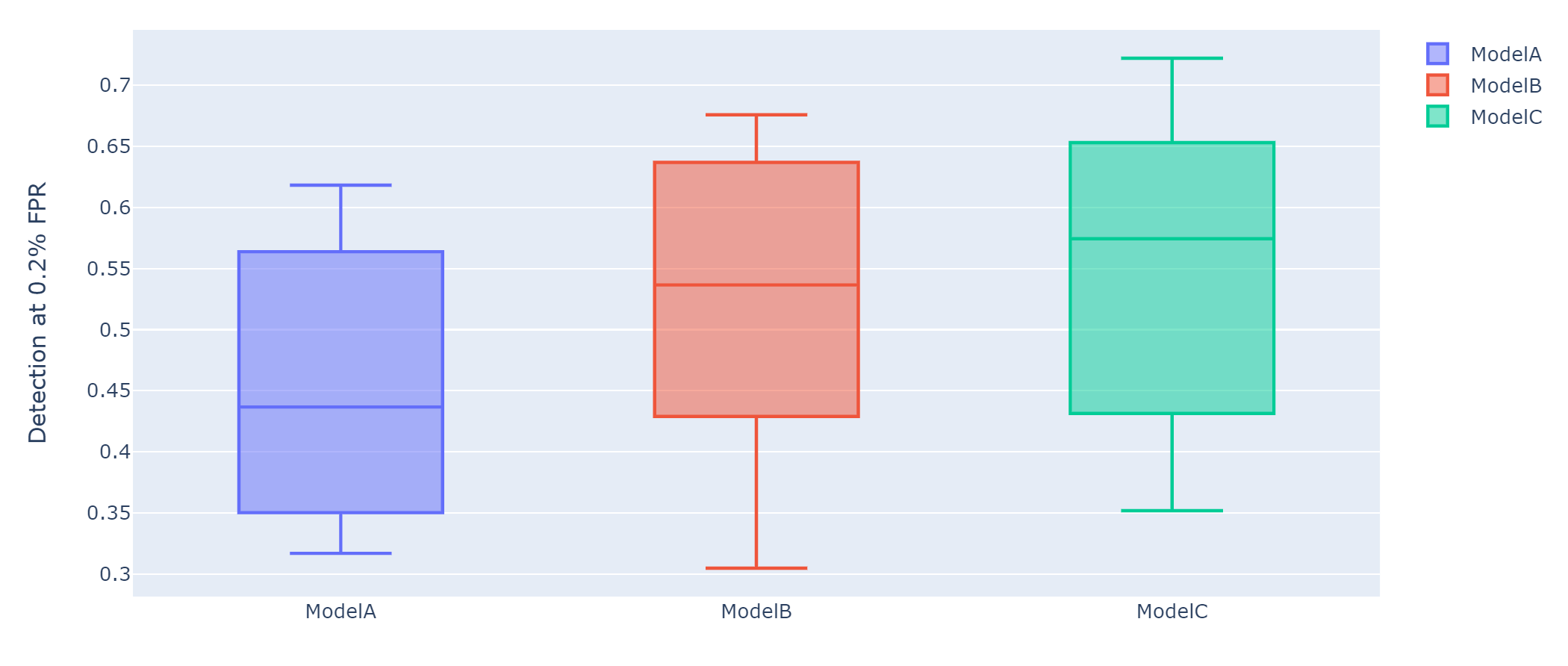}
        \caption{Detection at 0.2\% per model. All the models have a high variance, the more complex network ModelC achieves a better detection.}
    \end{subfigure}
\label{fig:errorbars}
\end{figure}

\subsection{Test results}

Each of our models in trained 10 times. We noticed that the variance is quite high between trainings and that taking the mean score is performing much better. From this point, except in Figure \ref{fig:errorbars} we will refer to the score of a model as the score of the ensemble.

In order to evaluate our algorithm, we compare our results to the antivirus in VirusTotal. As we built our dataset querying for files that have more than 10 detections, a small bias is created towards the antivirus but it still helps us understand how well our algorithm work compared to existing solutions.

We also compare the neural networks with a more classical Machine Learning-based approach for detecting malicious PDF. We trained a random forest on using the tags of the PDF files following  \cite{mai}. We selected the features using the TF-IDF method and tuned the parameters by taking the set that maximized the detection at 1\% FPR on the validation set.

\begin{figure}[t]
\centering
\caption{Bar chart of detection for all AV (in blue) and the proposed models on the test set (in green)}
\label{fig:bar1}
\centering \includegraphics[width=\textwidth]{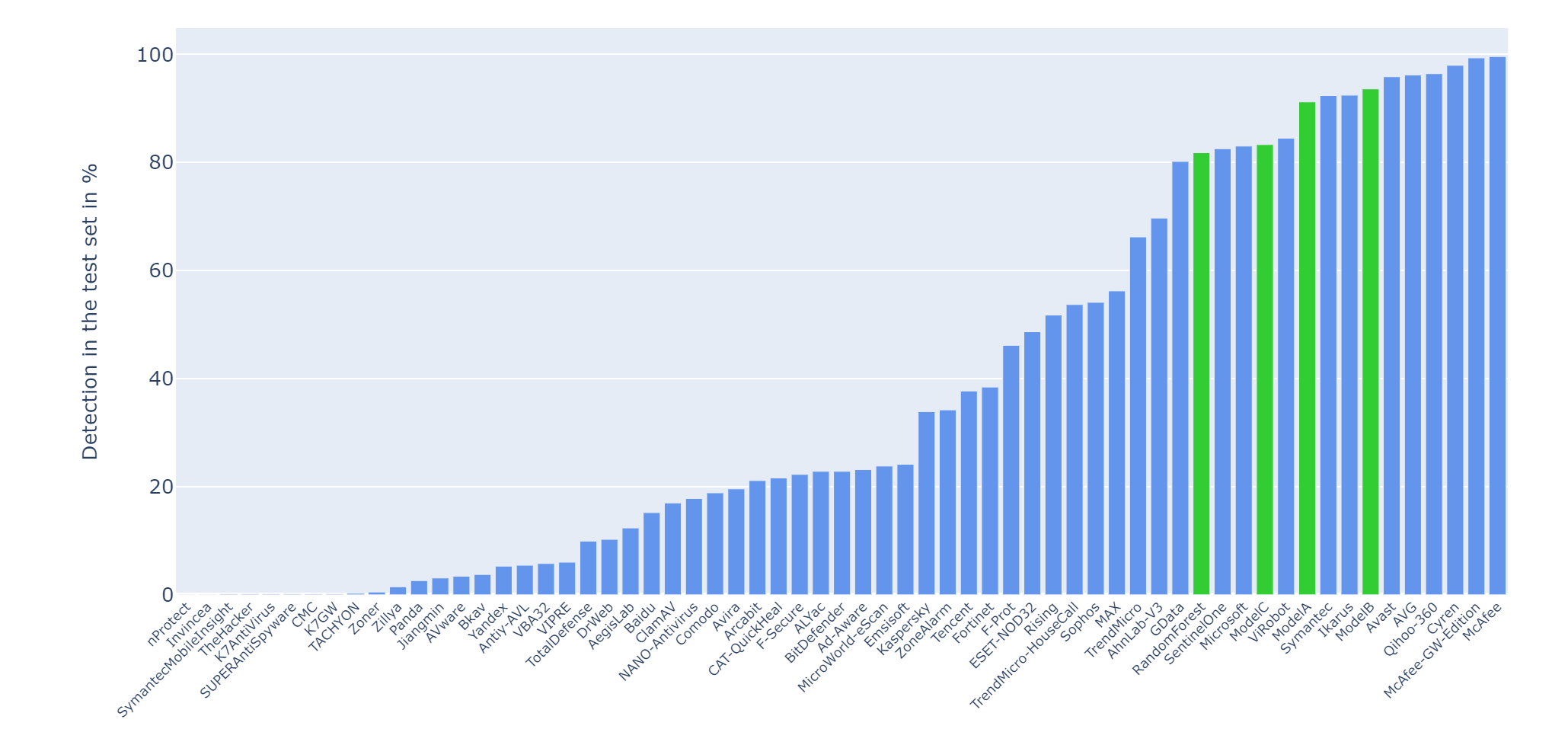}
\end{figure}

A ranking of all the algorithms together with the AV is provided in Figure~\ref{fig:bar1}. We used the detection at 0.5\% FPR as a reference for all the Machine Learning models.
The ROCs on train and test displayed in Figure \ref{fig:roc1}.
%
%Each of our models has been run 4 times, sampling with replacement on the train set. We show %the average result together with the interval on all the runs. 

%\begin{figure}[h]
%\centering
%\caption{Histogram of the AV detection}
%\label{fig:hist1}
%    \begin{subfigure}[b]{0.45\textwidth} 
%        \centering %\includegraphics[width=\textwidth]{AVdetectiontrain.png}
%        \caption{On the train set}
%    \end{subfigure}
%    ~ 
%    \begin{subfigure}[b]{0.45\textwidth}
%        \centering %\includegraphics[width=\textwidth]{AVModdetection.png}
%        \caption{On the test set}
%    \end{subfigure}
%\end{figure}

We can see that our best model ranks in the $7^{th}$ position of all the $52$ antiviruses available on the website at the time of the research.

It is worth mentioning here that most antiviruses use hand-crafted signatures and require significant human resources to accomplish this. In our model, no domain knowledge is required, and the signatures are learned automatically by the model.

As we can see in Figure \ref{fig:errorbars} both ModelA and ModelB have very high detection rates at 1\%FPR and 0.5\%FPR, with rather negligible differences in the detection rates.
For 0.2\%FPR ModelC outperforms the two others, although the variance is quite high for all the models. 
%[YM: This is referencing Figure 2, say it. Also maybe add the figure for 0.5\% and 0.2\%, or at least add a table with the numbers]
%
%We can also note that the ModelB is more robust and efficient than the two other, and maintain a high detection rate even at a FPR of $0.2\%$. Finally, we remark that ModelA still performs as well at 1\%FPR.
%
This suggests that imitating the AV work, with signature matching, seems like an effective approach for detecting PDF malware.

\subsection{Degradation with time}

One of the issues with using machine learning to detect malware is that our detection rate is expected to decrease with time. Indeed, malware are constantly evolving, as hackers keep refreshing their technics, so detection datasets become quickly outdated and models performance degrades on new samples.

In order to evaluate our robustness with time, we download a set of 9300 files from VirusTotal that were seen for the first time between the 15th of February and the 15th of March 2019. This makes a 5 months delay from the oldest file of our training set to this new set.
In this experiment, our detection for the ensemble of ModelB, using the same threshold of 1\% FP, is equal to $36\%$. We could have expected this drastic decrease as the malware are constantly evolving and the signatures have to be updated quite often.  

\begin{figure}[t]
\centering
\caption{Histogram of the AV detection}
\label{fig:hist1}
    \begin{subfigure}[b]{0.42\textwidth} 
    \caption{Score distribution for the test set and for the new files}
    \centering
    \includegraphics[width=\textwidth]{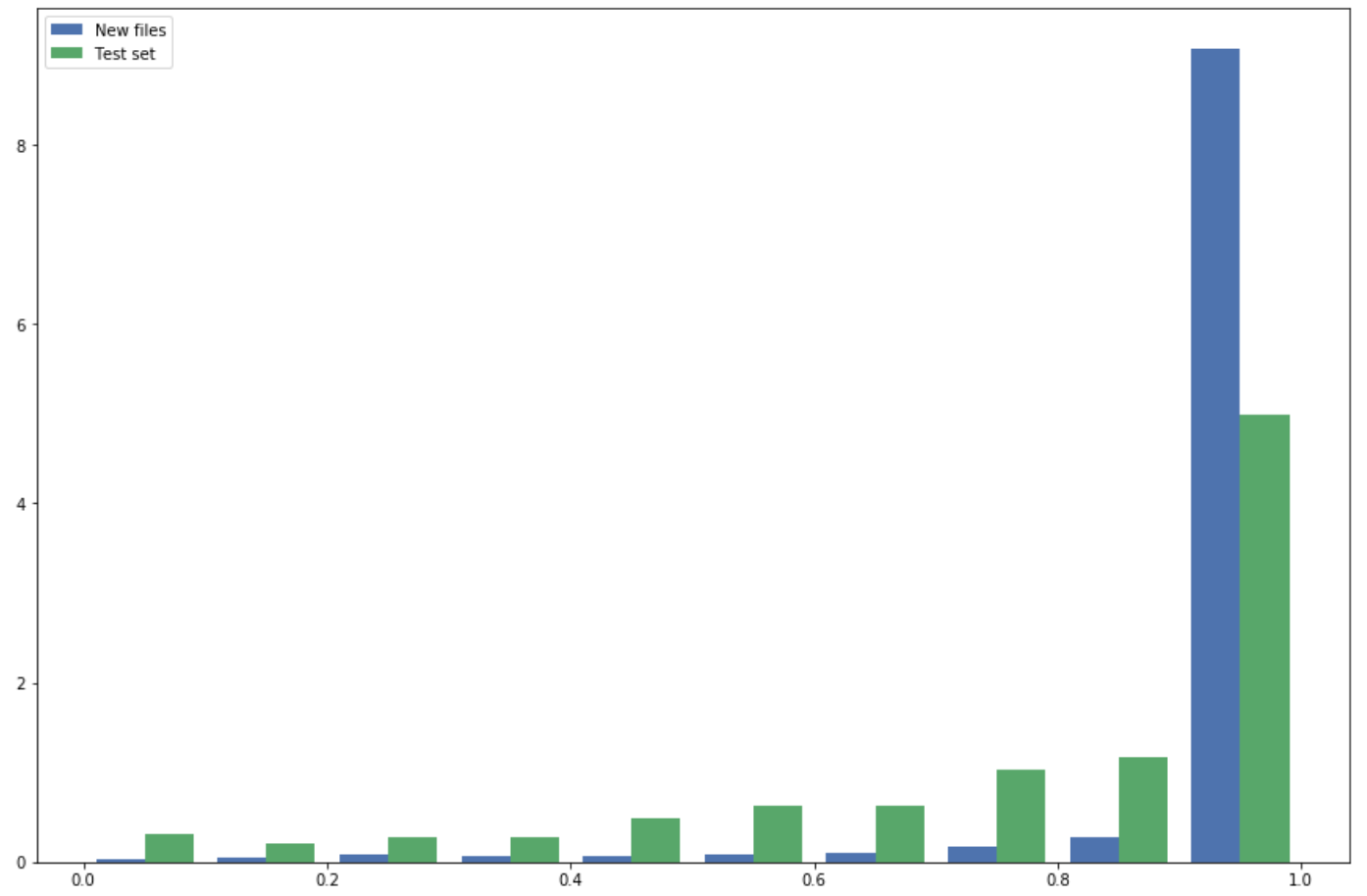}
    \label{fig2:scores}
    \end{subfigure}
    ~ 
    \begin{subfigure}[b]{0.43\textwidth}
    \caption{Detection of Sonbokli files two weeks after they were discovered}
    \centering
    \includegraphics[width=\textwidth]{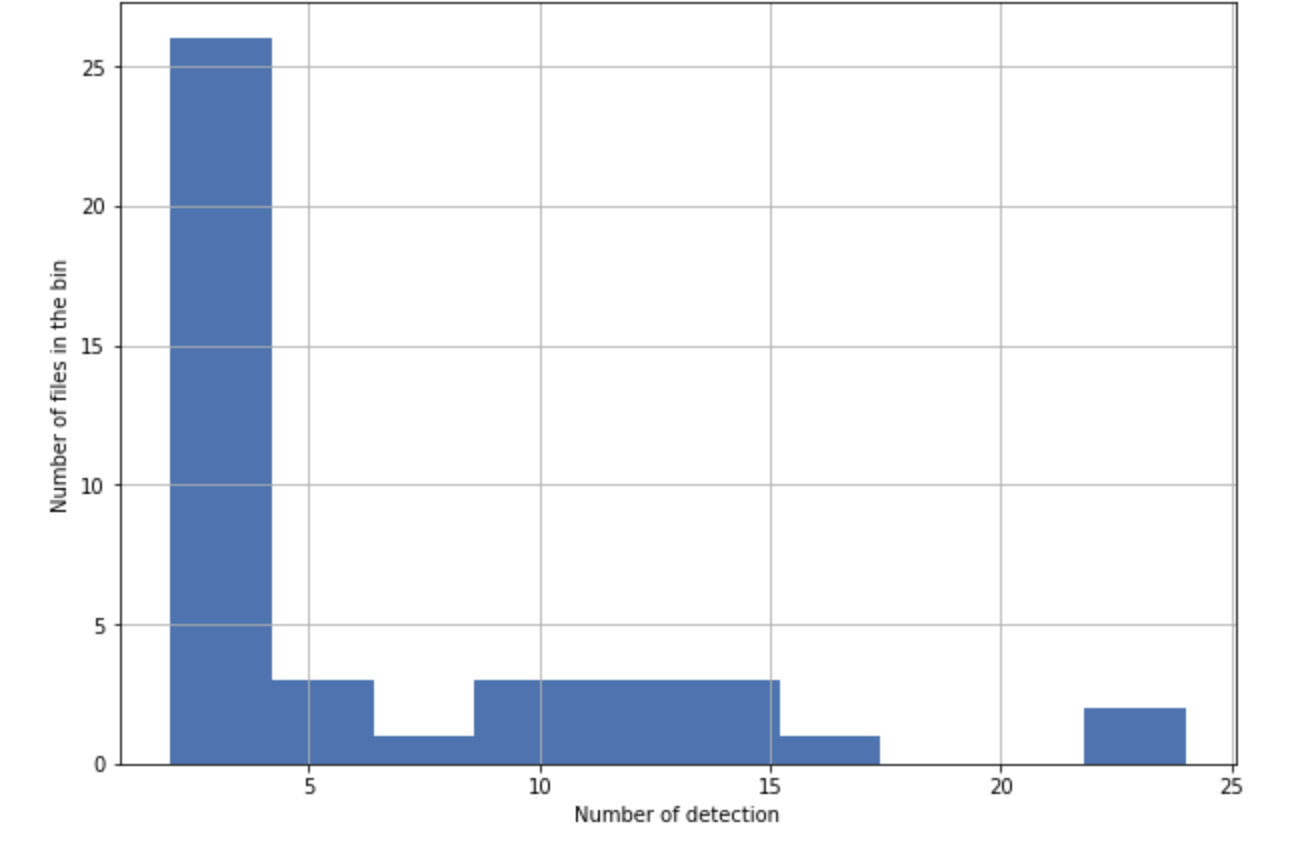}
    \label{fig2:sonbok1}
    \end{subfigure}
    \label{fig:2hist}
\end{figure}

\subsection{Zero days}
With the same goal of investigating the generalization capabilities of our model we also look for files that were almost undetected by antiviruses (AV) during our training phase and that started being increasingly detected later. Due to the limitations of the query language in VirusTotal we looked manually for some of those examples.

One file we found is labeled as belonging to the Sonbokli family by Microsoft. We downloaded 100 more PDF tagged with this name and having more than 10 antiviruses detecting them in July 2019. Surprisingly, most of these files are presenting the same kind of pattern in Antivirus detection. Initially, a very small number of vendors detect them on the first appearance (usually less than 4, all after January 2019), and, after some time, this number drastically increases, as more vendors detect them. 

We run our ensemble on those 100 files, which detected 74 of them. This means that our CNN was capable to learn a signature more robust than the one written by most antiviruses for this specific family of malware. This signature was generic enough to still be relevant 8 months after the end of the training period.

Finally, we look for Sonbokli files of the week of the experiment. We found 42 PDF files uploaded to VirusTotal between the 1st and the 7th of July 2019, none of them had more than 10 detections, and around half of them are detected by Microsoft only. Still, our network is able to detect 83\% of them as malicious. 
We show in Figure \ref{fig2:sonbok1} the number of antiviruses that detect each file, two weeks after we saw them for the first time.

\subsection{Contagio Dataset}

In the interest of evaluating our model in a more standard academical setup, we trained a CNN on the Contagio dump dataset. This dataset is widely used in previous research on Machine Learning applied to malicious PDF detection \cite{mai} \cite{smu}, \cite{cuan2018malware}. While  Cuan et al. \cite{cuan2018malware}, and Stavrou and Smutz \cite{smu}  are using all the files and evaluate their results using cross-validation, Maiorca et al. \cite{mai} are enriching with some benign files taken from the web (which represent 25\% of their overall set). Then, they used 6000 and malicious samples before running the experiment.
Note that the number of files used is always a bit different and no explicit list is provided in none of these papers.

We decide to run our ModelB on the Contagio dataset. As cross-validation can take a long time with a CNN we use the train/test/validation setup specified in \ref{sec:data}. 
We run our algorithm for 10 epochs without early stopping as we did not notice degradation on the results on the validation set during the training.

The results for all the models are specified in table \ref{table:contagio}.

We ended up obtaining a better accuracy than the one reported in the previous research. Because the way the experiment is done and the split is chosen is quite different among the researches, we cannot state that we outperformed previous algorithms, although it is fair to say that our results are comparable. We proved here that using CNN allows us to skip the tedious task of feature engineering when training an algorithm to detect malware in PDF files. 
\begin{table}[b]
\caption{Comparison of reported results on the Contagio Dataset for PDF files }
\begin{center}
 \begin{tabular}{||c l||} 
 \hline
 Model & Accuracy\\ [0.2ex] 
 \hline\hline
 \hline
 \textbf{CNN} & \textbf{99.83}\% \\
 \hline
 Maiorca et al. & 99.82\% \\
 \hline
 Smutz et al. & 99.81\% \\
 \hline
 Cuan et al. & 99.68\% \\
 \hline
 \end{tabular} 
\end{center}
\label{table:contagio}
\end{table} % Experiment 2

\section{Differentiating various families malware}
\label{sec:differentiating}

The main issue while trying to classify malware families is that creating their labels is much more complex. The main reason is that antiviruses' names are inconsistent and finding a good heuristic for labeling is a hard problem \cite{mag}. For that reason, we decide to use unsupervised learning to investigate how well our CNN can differentiate between different malware families.  We will use antivirus labels only as a reference for comparison.

In this experiment, we retrieve the 1231 malicious samples of our test set, then run a trained instance of ModelB on them while extracting the output of the global max pool layer.  This means that for each sample we will be represented by a vector in $\mathbb{R}^{128}$.

\subsection{Visualization}

We first start by displaying in 2D the vectors obtained after the global max pool layer. To do so we will use the common dimensionality reduction technique: T-SNE \cite{tsne}. We would like to compare our results with an antivirus' names. For that, we require the antivirus to detect the major part of our samples while having enough diversity in its labels.
The one we found that satisfies the best of both of those requirements is Microsoft. 

\begin{figure}[t]
\caption{T-SNE applied to the $2^{nd}$ layer}
\centering
\includegraphics[scale=0.4]{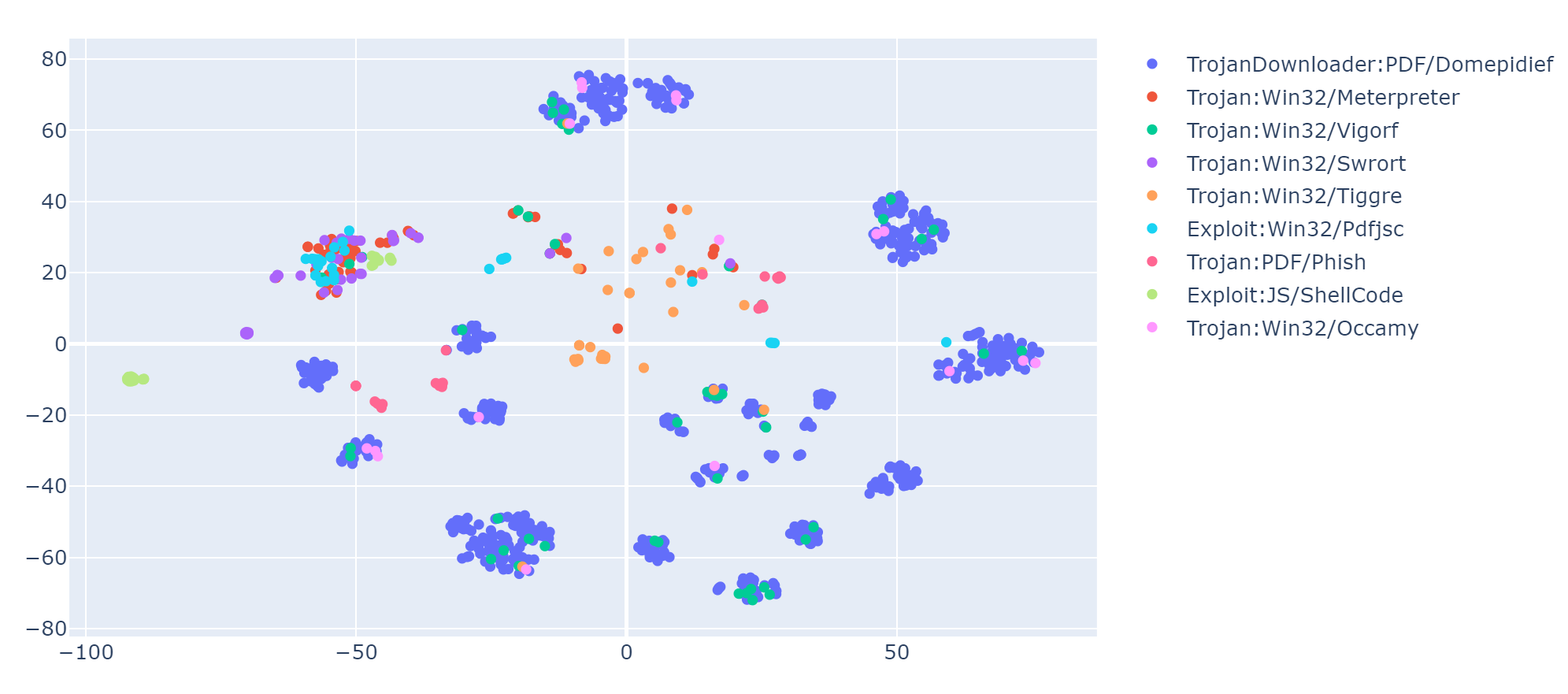}
\label{fig:tsne1}
\end{figure}

For convenience, we select only the 10 most common families, which drops only 10\% of the samples. We also remove the undetected, which represents around 7\% of our test.
The graph is presented in Figure \ref{fig:tsne1}.

We can easily distinguish the Phishing PDF (TrojanDownloader Domepidief), which is the most common family among the data and which is split among a few clusters which seem to be different variations of this malware family.  Another observation is that the families that are based on exploits end up pretty close on the graph, we can guess here that all those malware programs are exploiting the same kind of vulnerability on the system (Meterpreter, Swrort, Pdfjsc). By looking more specifically at some of these files in VirusTotal, we noticed that most of them are labeled ``js-embedded", indicating that they run Javascript on the host computer. This indeed is a strong indicator of maliciousness \cite{las} and explains why we see those groups next to each other.

\begin{figure}[t]
\centering
\caption{Homogeneity and Completeness by malicious files cluster}
    \begin{subfigure}[b]{0.45\textwidth} 
        \centering \includegraphics[width=\textwidth, height=4cm]{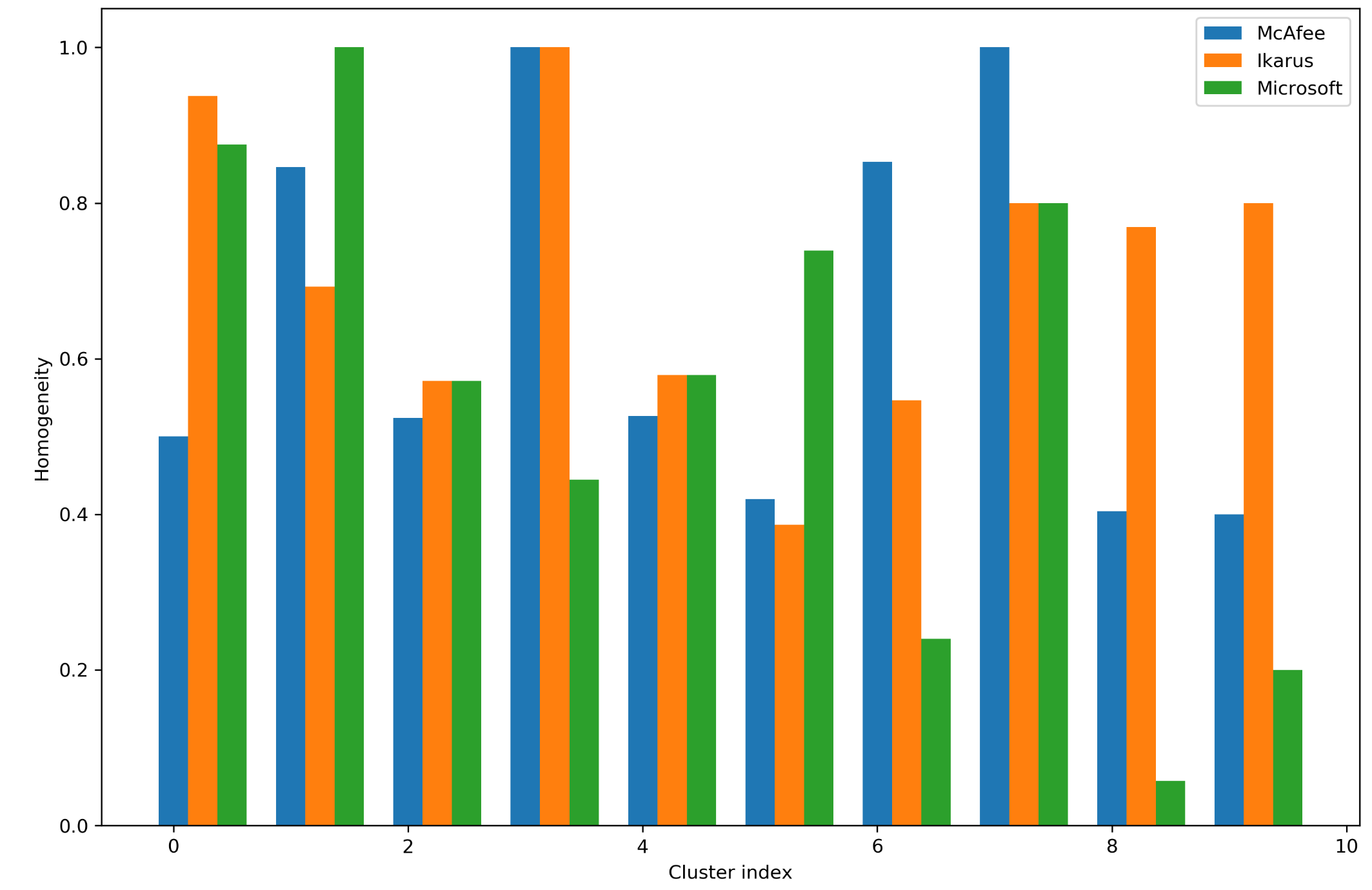}
        \caption{Homogeneity by cluster}
    \end{subfigure}
    ~ 
    \begin{subfigure}[b]{0.45\textwidth}
        \centering \includegraphics[width=\textwidth, height=4cm]{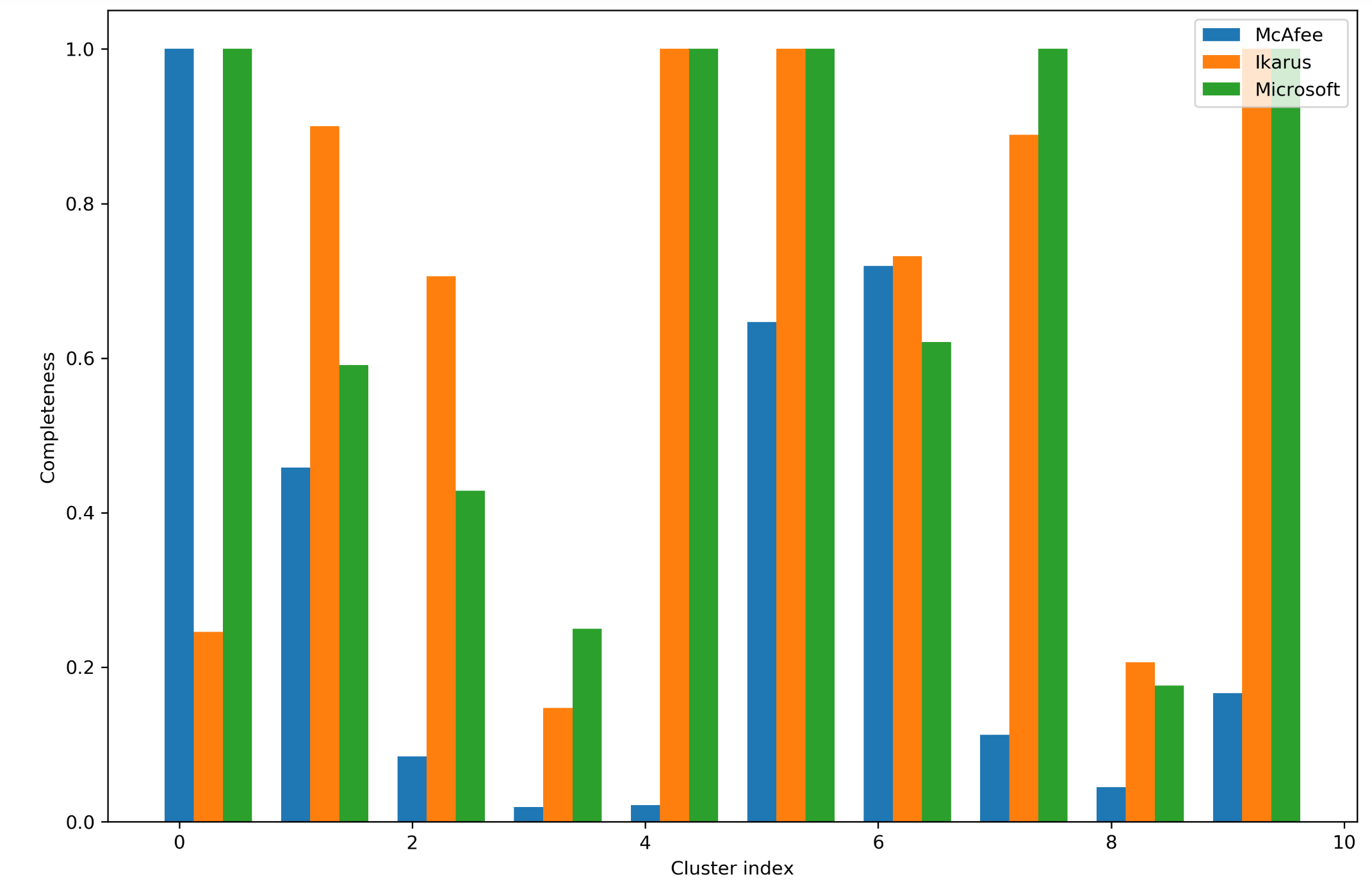}
        \caption{Completeness by cluster}
    \end{subfigure}
\label{fig:clustering1}
\end{figure}

\subsection{Clustering}

In order to dig deeper into the understanding of the network, we run a clustering algorithm on the vectors described above and compare the results with three antiviruses: Ikarus, Microsoft, and McAfee. We aim to evaluate two metrics: what part of each cluster has the same label (homogeneity) and how much of each family is contained in a specific cluster (completeness).

As a clustering algorithm, we decide to use the density-based algorithm HDBSCAN \cite{hdbscan}, as it does not require any hyper-parameter except the minimum of files in the cluster, does not make a prior assumption on the shape of the clusters and allows some files not to be grouped.
We run the algorithm on the samples without prior filtering, defining a minimum size of 10 files by cluster.

\begin{figure}[t]
\caption{Detection rate by clusters by AV}
\centering
\includegraphics[scale=0.23]{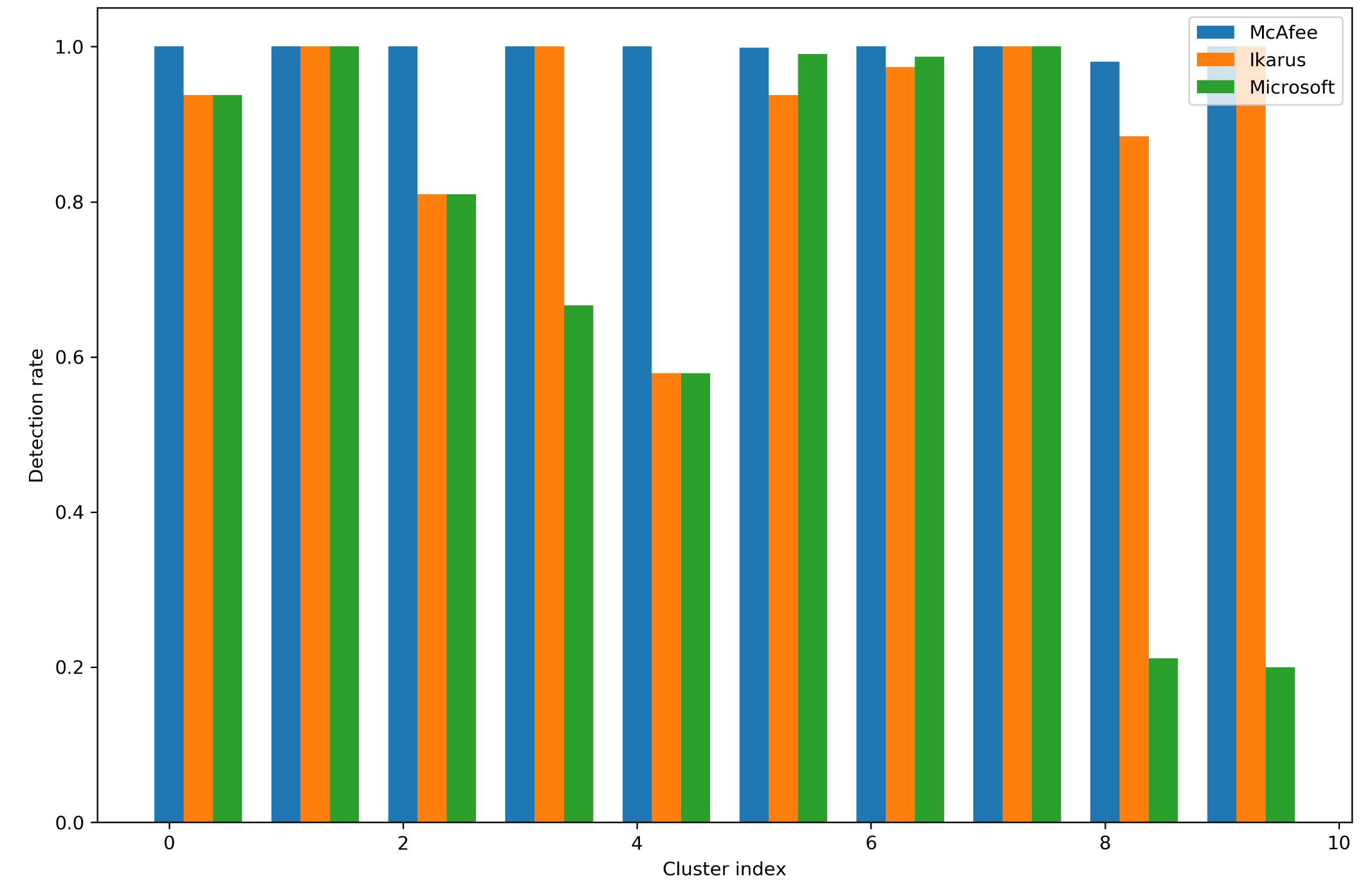}
\label{fig:detectionCluster}
\end{figure}

To make the results easily explainable we stick on a simplified definition of homogeneity and completeness. 
\begin{itemize}
  \item The homogeneity is estimated by the ratio between the most common label in the cluster with the number of elements in it.
  More formally, let $L$ be the set of all the names given by a specific antivirus. The homogeneity $H_i$ for cluster $C_i$ is given by 
  $$
  H_i= \max_{l \in L}{\frac{|\{(c,l):c\in C_i\}|}{ |C_i|}}
  $$
  \item The completeness is given by the ratio of the biggest family in the cluster that is contain in it. Let $\hat{l_i}=\argmax_{l \in L}|\{(c,\hat{l}):c\in C_i\}|$ the most common family in $C_i$. The completeness $T_i$ of $C_i$ is given by:
  $$
  T_i=\frac{|\{(c,\hat{l_i}):c\in C_i\}|}{|\cup_j\{(c,\hat{l_i}):c\in C_j\}|}
  $$
\end{itemize}
Note that, for the computation of the homogeneity we do not take into account the files in the clusters that are not detected.  The detection rate by cluster and by antivirus can be found in Figure \ref{fig:detectionCluster}. 
%[YM: why not have it in the main text, we have no page limit]

Figure \ref{fig:clustering1} is displaying our results by clusters and by antivirus for the two introduced metrics.
Out of the 1231 files, only 282 did not end up in any of the 10 created clusters.
The most common family in each cluster are listed in the table \ref{table:clusterNames}. 
%[YM: why not have it in the main text, we have no page limit]

\begin{table}[b]
\caption{Most common family name per cluster }
\begin{center}
\resizebox{\textwidth}{!}{%
 \begin{tabular}{||c l l l||} 
 \hline
 Cluster & Microsoft &  Ikarus & McAfee \\ [0.2ex] 
       
 \hline\hline
 0 & Exploit:Win32/Pdfdrop.D & possible-Threat.PDF.Acmd & Suspicious-PDF.gen.a\\ 
 \hline
 1 & Exploit:JS/ShellCode.gen & Exploit.PDF-JS &  RDN/suspicious-pdf.gen\\
\hline
 2 & Trojan:Win32/Tiggre!plock & Trojan.SuspectCRC & RDN/Generic.dx \\
 \hline
 3 & Trojan:Win32/Meterpreter.O & possible-Threat.PDF.Acmd & Artemis \\ 
 \hline 
 4 & Exploit:Win32/CVE-2012-4914 & Exploit.Win32.CVE-2012-4914 & Artemis \\ 
 \hline
 5 & ...PDF/Domepidief.A & ...PDF.Domepidief &  Artemis  \\
 \hline
 6 & Exploit:Win32/Pdfjsc & PDF.Exploit.PDF-JS & Exploit-PDF.bk.gen \\
 \hline
 7 & Exploit:SWF/CVE-2010-1297.D & Trojan.Script &  Exploit-PDF.bk.gen\\
 \hline
 8 & Trojan:PDF/Sonbokli.A!cl & Trojan.PDF.Phishing & Artemis \\
  \hline
 9 & Trojan:HTML/Brocoiner.N!lib & Trojan.JS.Agent & RDN/Generic Downloader.x \\
  \hline
 \end{tabular} %
 }
\end{center}
\label{table:clusterNames}
\end{table}

\subsection{Analysis}

We can observe a few different types of clusters. First, clusters that match well at least one antiviruses' prediction on a family. Clusters 1 and 7 are catching well a specific exploit and cluster 0 PDF droppers. Cluster 3 is homogeneous but not complete, it identifies a specific variant that antivirus classify to broader families. This indicates that our algorithm does find a specific identifier for these malware while antiviruses designate them with generic names.
Some other clusters are perfectly complete according to one or more vendors but not homogeneous. This means that our algorithm is generating a signature that catches a few families together. This is the case for the exploit cluster 4, cluster 5 containing mainly phishing files and cluster 9.
It worth remarking the difference between antivirus at naming families. For example in clusters 0 and 7 Microsoft finds a specific name for the malware that we don't find anywhere else, while Ikarus uses a generic designation in cluster 0 and McAfee in cluster 7.
While having the highest detection rate, McAfee tends to use less specific names than the two other, this is reflected by a higher homogeneity and a smaller completeness.
 % Results and Discussion

\section{Conclusion}
\label{sec:conclusion}

We introduced a Convolutional Neural Network to detects malicious PDF using only the byte level of the PDF files. We have shown that our network achieves a performance comparable to some of the best antivirus in the market without requiring any preprocessing or feature extraction. It is able to detect files discovered a few months after its training period and malware that are undetected by most antivirus. Finally, the network proved to be capable of distinguishing accurately various malware families. %\newline

Using this kind of approach could save significant time for malware analysts and automate detection. It will probably be one of the solutions chosen by antivirus companies to cover the drastic increase of new malware files discovered every day.

%\newline
In future work, we hope to investigate additional neural network architectures including Recurrent Neural Network (RNN). Additional future research is considering additional  meaningful
embedding for the raw PDF files, for example, by including some metadata information like tags or xref tables. % Conclusion

\newpage

\label{Bibliography}
  % Change the left side page header to "Bibliography"
\bibliographystyle{abbrv}
\bibliography{Bibliography}  % The references (bibliography) information are stored in the file named "Bibliography.bib"

\newpage
\appendix % Cue to tell LaTeX that the following 'chapters' are Appendices

\section{Networks architectures}
\label{appendix:networks}

\begin{figure}[h]
\caption{Model A}
\includegraphics[width=\linewidth]{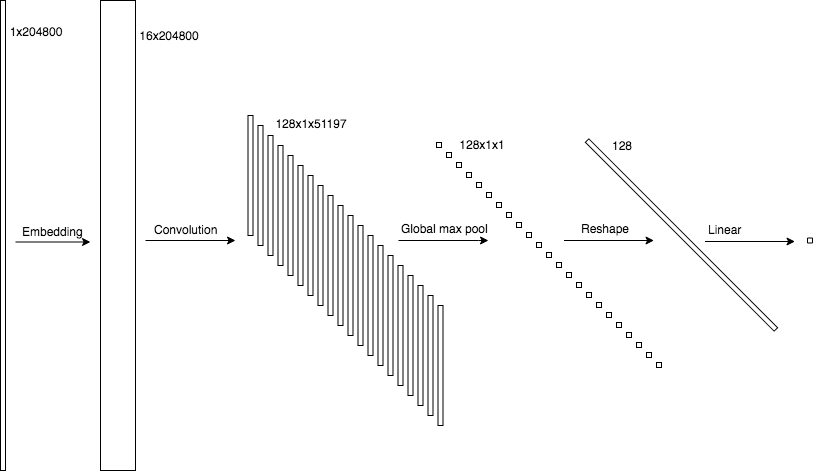}
\end{figure}

\begin{figure}[h]
\caption{Model B}
\includegraphics[width=\linewidth]{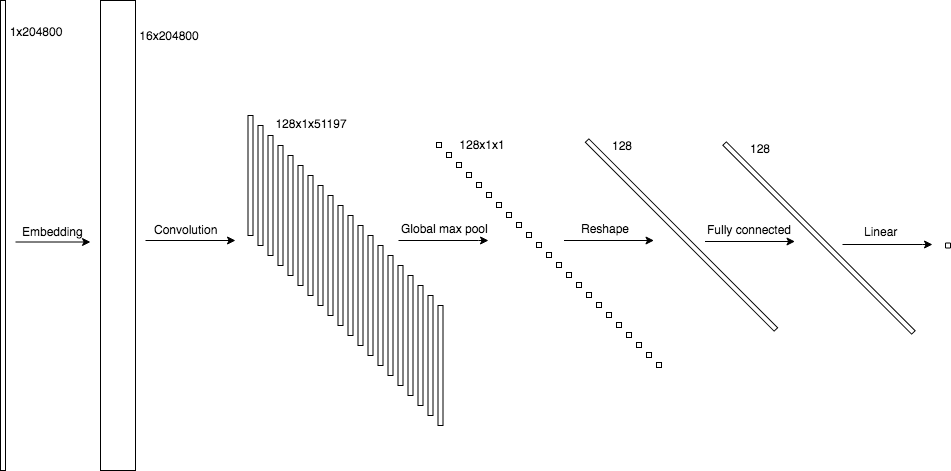}
\end{figure}

\begin{figure}[h]
\caption{Model C}
\includegraphics[width=\linewidth]{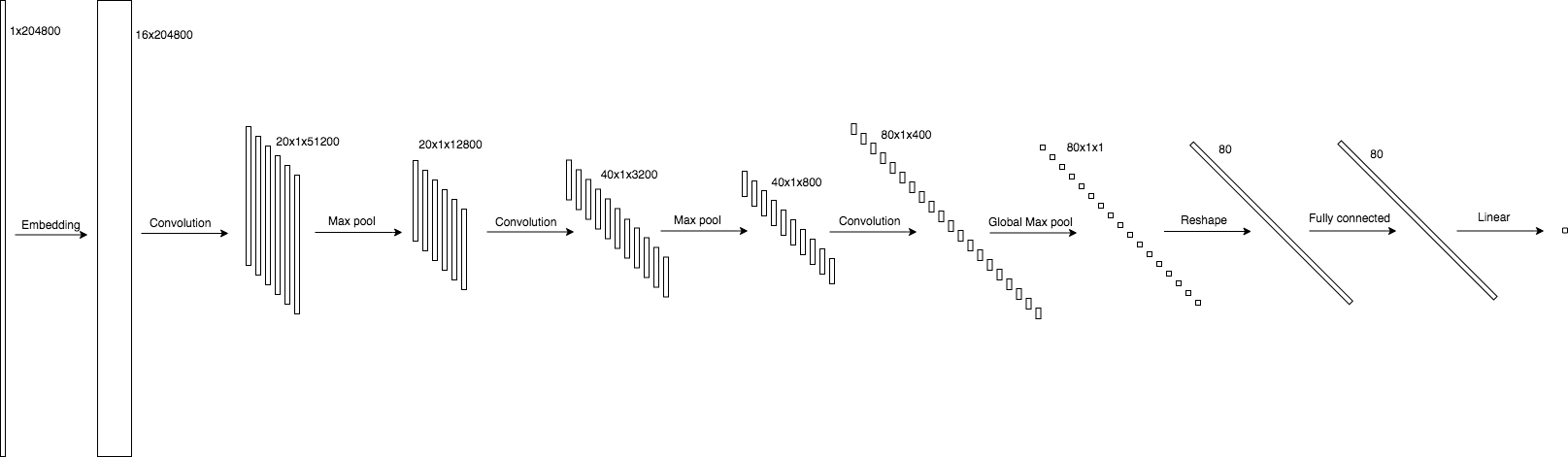}
\end{figure}\textbf{}

\end{document}